\magnification=1200
\parindent=0pt
\parskip=8pt
\def\pp{\noindent\parshape 2 0truecm 15truecm 2truecm 13truecm}
\leftline{\it To appear as a Review Article in Nature}
\bigskip
\centerline{\bf The Formation and Evolution of Galaxies}
\bigskip
\centerline{\bf Richard Ellis}
\centerline{\bf Institute of Astronomy, Cambridge, UK}
\vskip 0.5truein
{\bf Galaxies represent the visible fabric of the Universe and there has
been considerable progress recently in both observational and theoretical
studies. The underlying goal is to understand the present-day diversity
of galaxy forms, masses and luminosities. Popular models predict the bulk
of the population assembled recently, in apparent agreement with optical
observations.  However, numerous uncertainties remain, including the
role that dust might play in obscuring star-forming systems. Astronomers
now seek more detailed tests to verify that the Hubble sequence of types
arises from transformations driven by the dynamical assembly of smaller
systems. Multi-wavelength surveys and studies of the resolved internal
properties of distant galaxies promise answers to these fundamental
questions.}

\bigskip

The name Edwin Hubble has become synonymous with progress in understanding
galaxies. Hubble opened the extragalactic era by demonstrating that
galaxies are stellar systems outside our own Milky Way$^1$. His name
now graces a most successful space observatory which has, since its
refurbishment in 1993, worked in harmony with ground-based telescopes to
transform our view of the distant Universe. An important goal of the 
Hubble Space Telescope (HST) is to directly establish the evolutionary 
history of normal galaxies. HST offers resolved data enabling us to 
perform detailed astrophysical studies on the internal properties of 
remote galaxies. Together with independent progress made in understanding 
how structure forms from initial density fluctuations soon after the 
Big Bang, a picture is emerging on how galaxies form and evolve.

A major question we hope to address is the origin of Hubble's
morphological sequence$^2$. Is there any physical basis for this
taxonomic scheme? Hubble noted a strong correlation between the
relative prominence of the central `bulge' and `disk' components along
his sequence. Bulge components (including ellipticals) are generally
redder than spiral disks and contain less gas, which can be interpreted
via the early rate of conversion of gas to stars.  Ellipticals could
have formed their stars during a rapid collapse leading to present-day
stellar populations which are uniformly old and red. By contrast,
spirals suffered a more extended star formation history retaining gas
and young stars to the present epoch$^3$. These different `metabolic
rates' might reflect the physical nature of the initial assembly. If
primordial gas clouds had time to collide and dissipate energy, a large
collapse factor would be permitted. Angular rotation associated
with the primordial cloud would amplify during collapse leading to a
rapidly-rotating disk. Since ellipticals are slowly-rotating, their
collapse factor would then have been modest and their high mass densities
might therefore reflect a much earlier period of formation$^{4}$.

For many years, most astronomers believed in a single `epoch of galaxy formation'$^5$ and sought to find `primaeval galaxies' e.g. the 
luminous star-forming precursors of present-day ellipticals. 
The absence of examples in deep emission line searches$^6$ and faint 
redshift surveys was the first hint that galaxy formation might be a 
gradual process rather than one confined to a narrow period in time.  
Moreover, perhaps galaxies did not evolve as isolated objects.
Local galaxies interact at a rate which is likely to have been higher in 
the past. Numerical simulations indicate gas-rich spirals, when they 
collide, can produce smooth components which look like bulges and
ellipticals$^7$. Upon closer examination, local bulges and ellipticals 
reveal a bewildering diversity. Some ellipticals have gaseous components 
and harbour small disks, many bulges are relatively blue. The enigmatic
lenticular or `S0' galaxies have disks and bulges, but without active 
star formation or spiral arms. Such observations may imply that disk systems 
were predominant at early times and that subsequent processes transformed 
some to give the wider range of types seen today. These conjectures can 
be tested by examining the morphology of galaxy populations viewed at 
earlier times.

\smallskip
{\bf Understanding the Growth of Structure}
\medskip

The most popular model accounting for the growth of structure is the
cold dark matter (CDM) model$^8$ which can reproduce a growing body
of observations including the angular power spectrum of fluctuations
in the microwave background and the large scale distribution of local
galaxies$^9$. Its predictive power is well-developed due to the enthusiasm
and foresight of its advocates$^{10,11}$. However, it is conceivable that
alternatives theories might still be developed which could also match
the observations. We should aim to check the physical principles of a
model rather than be content solely with the fact it reproduces many
of the observations.

CDM theory posits that a large fraction of cosmic matter is dark and 
non-baryonic$^{12}$. Non-baryonic material has the advantage of decoupling 
from the cosmic plasma at early times with a characteristic spectrum 
of density fluctuations that merge hierarchically providing early seeds 
for later galaxy formation.  A young galaxy of gas and stars is embedded 
in a more massive dark matter halo and its evolution is governed by the 
nature of that underlying halo. Baryonic gas accretes gravitationally onto 
the halos and is shock-heated. Stars can only form when gas has cooled
and this is hindered by various `feedback' processes.

The cooling timescale for primordial gas clouds provides a natural
explanation for the physical sizes of present-day galaxies$^{13.14}$.
If the cooling time exceeds that for dynamical collapse, a cloud can
readjust its temperature and collapse quasi-statically. However, if
cooling is too rapid, then virial equilibrium is never reached and
rapid collapse and fragmentation ensues. Such considerations imply that
the onset of star formation in a young galaxy is a gradual process
governed by gas cooling and feedback as well as the rate at which mass
is assembled. For CDM, the latter is fixed by the need to match the
observed abundance of present-day rich clusters which determines the
epoch at which structure on any mass scale forms. A long-standing
prediction$^{15}$ is that the bulk of the star formation might have
occurred as recently as a redshift $z\simeq$1.

The appeal of the CDM picture lies in its ability to broadly match many
of the observations of large scale structure. Whereas puzzles remain,
these data sample the underlying density field and so relatively few
model parameters are involved. However, in reproducing detailed
observations of galaxies, a larger suite of astrophysical parameters is
needed including the physical details of star formation and feedback
and dynamical details of how galaxies subsequently merge to form larger
morphological structures. The philosophy adopted has been to choose
parameters such that the present-day distribution of galaxy properties
is reproduced. A comparison of CDM predictions of how galaxies evolved 
to their present forms with distant data is therefore a crucial test.

\smallskip
{\bf Recent Observational Progress}
\medskip
 
The refurbished HST and surveys with ground-based large telescopes
equipped with efficient faint object instrumentation has led to
breathtaking progress. Key developments point to a remarkably recent
era of galaxy formation. The results remains quantitatively uncertain and
do not, by themselves, provide any theoretical insight into the processes
involved. However, the empirical picture emerging is qualitatively
similar to that predicted by CDM.

Until recently, the vanguard of progress has been via statistical studies 
at optical and near-infrared wavelengths. A remarkably simple cosmological 
probe of the early history of star formation is the logarithmic slope,
$\gamma$=$dlog\,N/dm$, of the galaxy number-apparent magnitude counts. 
To yield a finite value for the extragalactic background light, $\gamma$ 
must ultimately drop below 0.4. Deep ground-based imaging 
surveys$^{16,17}$ presented evidence for a change in slope at $B\simeq$25 and 
this is now confirmed from deeper counts with HST$^{18,19}$ (Figure 1). 
A similar effect is seen at near-infrared wavelengths$^{20,21}$ suggesting 
a transition at some redshift with the bulk of the fainter sources being 
drawn from intrinsically-less luminous sources at similar redshifts.
Galaxies at the point of inflexion must provide a dominant
contribution to the integrated background light.

Deep spectroscopic surveys conducted with ground-based telescopes
equipped with multi-object spectrographs have been an indispensible
tool in cosmology since the mid-1980's. Their main task has been to 
determine the redshift distribution of statistically-complete samples 
of galaxies at various magnitude limits$^{22-26}$. Such surveys have 
almost reached the inflexion in the optical counts and confirm a 
surprisingly low mean redshift $z\simeq$0.8-1 at faint limits. 
Few $B$=25 sources are distant luminous young galaxies as expected for 
a narrowly-defined epoch of galaxy formation.

These surveys also indicate that the steeper count slope at brighter
apparent magnitudes is a strong evolutionary effect representing a
significant increase to redshifts $z\simeq$1 in a population of star
forming galaxies with blue colours$^{22}$, intense emission lines$^{23}$
and irregular HST morphology$^{27,28}$. The combination of HST morphology
and deep spectroscopy has been particularly effective in pointing to an
era of intense star formation at z$\simeq$1-1.5$^{29}$. Some component
of the recent decline in star formation may be associated with the slow
demise in regular spirals$^{30,31}$ but a surprisingly large evolutionary 
signal arises from low mass irregulars whose local counterparts 
are unclear$^{19}$.

There are not yet any systematic surveys beyond $z\simeq$1 whose 
selection is based purely on apparent magnitude. However, multi-colour 
imaging can be used to isolate distant source by detecting the Lyman
limit$^{32}$. Absorption by neutral hydrogen occurring in the intergalactic
medium and within galaxies produces a natural cosmic signature: a sharp 
drop in flux when the absorption edge is redshifted into the observer's 
frame. The technique is simple and effective. A galaxy `drops out' of 
view in a particular filter if the Lyman limit (912\AA\ ) has been 
redshifted beyond that wavelength (Figure 2). An early application$^{33}$
suggested that most $R$=25 sources lie below $z$=3 and that such 
`drop-outs' were rare but a convincing demonstration awaited confirmatory 
Keck spectroscopy of candidates selected by Steidel, Lowenthal and 
their colleagues$^{34-36}$.

A considerable amount is now known about these `Lyman limit galaxies' including
their spectral properties$^{37}$, physical sizes, HST morphologies$^{38}$ and
clustering properties$^{39}$. Of greatest importance here is the 
rest-frame ultraviolet luminosity distribution$^{40}$ which yields an estimate
of the comoving density of massive star formation at early times. The 
activity at $z>$2 appears to be significantly less than that observed 
in the all-inclusive redshift surveys at $z\simeq$1, consistent with the 
simple interpretation of the number counts discussed earlier. Moreover, 
by extending the technique to longer wavelength filters, the contribution 
from Lyman limit galaxies at higher redshift can be found. A further decline 
in activity is seen from $z\simeq$2.3 to $z\simeq$4$^{41}$.

The history delineated by these low and high redshift surveys suggests a 
peak in star formatgion between $z\simeq$1.2 and 2.5 where, currently,
there is no complete spectroscopic sample. This `gap' in coverage 
arises because the Lyman limit technique can currently only be located with 
optical facilities (including HST). Magnitude-limited spectroscopic 
surveys within 1.2$<z<$2.5 are similarly challenged by the need to follow 
diagnostic features in the near-infrared where spectrographs have to battle 
an intense sky background. Our limited knowledge of sources in this 
interesting region is provided from methods based on multi-band
photometry$^{42}$.

\smallskip
{\bf Is the Observational Picture Correct?}
\medskip

The star formation history delineated above has attracted considerable 
attention both by CDM theorists who claim vindication of their 
predictions$^{15,43}$ and by skeptical observers who have set out to
demonstrate it incorrect$^{55,58}$! According to CDM, the Lyman limit 
galaxies are precursors of massive cluster galaxies whose assembly 
history has been accelerated in the associated density peaks; the 
bulk of normal field galaxies are put together later as observed$^{44}$. 

The morphological appearance of the faintest sources gives qualitative
support to the idea that galaxies at z$\simeq$1-3 are still assembling.
Their mean physical size is small notwithstanding redshift-dependent
biases$^{45,46}$ and a high proportion are irregular with evidence of
ongoing merging$^{47}$. Deep infrared redshift surveys$^{24}$ sensitive
to stellar mass rather than young stars find a paucity of luminous
z$>$1 sources as expected if galaxies assembled late$^{48,49}$.
Independent evidence for recent galaxy formation (not reviewed here -
see ref. 50) arises from the low metallicity$^{51-53}$ and high gas
content$^{54}$ of material observed in absorption to high
redshift quasars although such interpretations depend critically on
whether the available data reliably serve as representative probes of
the global composition.

The star formation history will inevitably be refined quantitatively
as larger surveys are conducted but a decline beyond $z\simeq$2 would
represent a profound result for cosmology; less than 20\% of the observed
star formation would have occurred before that epoch. Recent discussions
have centred on two main criticisms. 

Firstly, it is possible that these optical datasets are being
misinterpreted.  The flux emerging from a Lyman limit galaxy will be
attenuated significantly at ultraviolet wavelengths by dust clouds and
thus the true star formation rate in these sources may be
underestimated. The decline in activity beyond $z\simeq$2 could be
entirely due to dust extinction. This cannot be addressed without
making some assumptions about the intrinsic properties of the Lyman
limit galaxies. If their ultraviolet properties are similar to local
star forming galaxies, dust correction leads to an upward correction to
the high z star formation rates of between 3$^{37,40}$ and 15$^{55}$. A
more extreme model where half of the present-day stars formed at
$z>$2.5 and were shrouded in view can be made consistent with the
global history of light at various wavelengths but could overpredict
the metal mass density observed in high z absorbers$^{56}$.
 
A more serious possibility is that the optical samples are incomplete
probes of the star forming population. Nearby starbursts output a significant
fraction of their radiant energy in the far infrared region via thermal
emission from dust clouds heated by shrouded young stars$^{57}$. Placed at
high redshift they could be missed altogether by the optical observers. 
Satellites such as the Infrared Space Observatory (ISO) are designed to
be sensitive to such radiation but at the expense of a poor angular
resolution$^{58}$. A moderate surface density of thermally-emitting sources 
has been found in deep exposures of the HDF but the optically-based star
formation history would only be significantly distorted if these sources 
were very distant. 

A promising way to address this problem follows the commissioning 
of SCUBA, a ground-based sub-mm array detector on the James Clerk 
Maxwell Telescope which has a better astrometric precision. Sub-mm 
receivers are very effective in searching for distant dusty 
sources because the redshifted thermal spectrum yields a detection
efficiency that hardly changes with redshift$^{59}$. The first genuine sub-mm
source counts are now emerging$^{60-62}$ and spectroscopic identifications
are available for a few sources$^{63}$(Eales et al, in prepn.) 

The change to the optically-derived star formation history would be
greatest if the bulk of the SCUBA sources lay beyond $z\simeq$2. The
volume-limited nature of the sub-mm surveys makes a powerful point.  A
dusty star-forming galaxy could be viewed to z=10 to the flux limits
now being achieved with SCUBA yet, where optical follow-up is
available, a high fraction of the sources have optical detections
and/or spectroscopic redshifts with $z<$3.5$^{62,63}$.

The quantitative form of the global star formation history (Figure 2)
is likely to change, both from improved optical data and more extensive
sub-mm surveys. One should not underestimate, however, the large
uncertainties in converting the sub-mm fluxes into star formation rates.
Conceivably the peak in activity at z$\simeq$1-2 will shift to higher 
redshift. However, the detailed shape of this curve is less important 
in testing models than the question of whether there is a sizeable 
population of massive star-forming galaxies at $z>$3. At the time of 
writing, there is no convincing evidence to challenge the notion that 
galaxy assembly occurred gradually over 1$<z<$3 as predicted by CDM.
 
\smallskip
{\bf Testing Hierarchical Models for Galaxy Formation}
\medskip

To what extent does a similarity between the predicted star formation
history and that observed imply support for the hierarchical models? The global
history integrates over the evolutionary behaviour of galaxies of all 
luminosities and morphologies thereby hiding many details. More precise 
tests must take advantage of the resolved properties of distant galaxies 
available with HST to probe directly the physical processes that lead to 
the diversity of the present day galaxy population. 
 
The dissipative collapse of primordial gas clouds in the CDM picture 
leads to rotating disk systems. External factors govern the structural 
and morphological evolution$^{64}$. Disks formed at high redshift slowly
merge providing a continuous supply of newly-born ellipticals 
and lenticulars.  

One of the most fundamental observations relating to the origin of
the Hubble sequence is the `morphology-density relation'$^{65}$ (Figure 3).
Ellipticals are predominantly found in cluster environments and noticeably
scarce in the low density `field'. Does this indicate that ellipticals
were produced at high redshift only in those regions destined to become
clusters (e.g. via a rapid initial conversion of gas into stars), or 
can ellipticals be produced subsequently through the mergers of spirals
as postulated by CDM?

Early ground-based work$^{66}$ tracked a rise with redshift in the 
fraction of star-forming galaxies in rich clusters and HST images show 
this is accompanied by a radical shift in galaxy morphologies$^{67,68}$.
By $z\simeq$0.5 spirals are almost as frequent in proportion in dense 
cluster cores as they are in the field. A particularly striking result 
is the rapid increase with time in proportion of lenticular (or `S0') 
galaxies, strongly suggesting they are transformed spirals.

Hypotheses advanced for the remarkably recent demise of cluster spirals
include dynamical friction in the cluster potential$^{69}$, gas
stripping of infalling field galaxies by a dense intracluster
medium$^{70}$ and galaxy-galaxy merging$^{71}$ induced perhaps by the
hierarchical assembly of clusters. Strong radial gradients observed in
diagnostic spectral features in distant clusters$^{72,73}$ support the
idea that gas-rich field galaxies suffer truncated star formation as
they enter the cluster on radial orbits. Although these transformations
are not yet understood, the observations demonstrate that recent
environmental processes shaped the morphology-density relation urging
us to take seriously the suggestion that morphology may be the result
of dynamical processes including ones associated with merging of dark
matter halos.

The important test objects here are the elliptical galaxies.
Traditionally their uniform red colours and smooth isophotes of
ellipticals indicate a rapid collapse at high redshift$^{74,4}$.  An
appealing argument in support of this picture is that homogeneous
populations of old cluster ellipticals can be found at high
redshift$^{75,76}$ suggesting that at least some completed their star
formation before $z\simeq$3. However, in view of the accelerated
evolution expected in dense environments$^{44}$, a more critical test
would be to examine the age distribution of field ellipticals.

Unlike clusters where the majority of sources are being seen at a 
single epoch, redshifts are necessary for each object in a field sample. 
Using optical colours to select candidate ellipticals from 
ground-based redshift surveys$^{22}$, Kauffmann et al$^{77}$ claim 
only a third of the present day comoving density of ellipticals at 
$z\simeq$0.8 are following evolutionary paths consistent with long-lived
stellar populations formed at $z>$3. Such a rapid decline in established 
red field ellipticals is consistent with calculations which equate 
morphology with the bulge/disk ratio produced in a dynamical merger$^{78,64}$. 

Such dramatic evolutionary trends should be easy to test with HST
data.  Through a study of areas for which extensive ground-based
spectroscopy had already been completed$^{22,23}$, HST images are now
available for over 300 field galaxies of known redshift$^{30}$. A
significant scatter is seen in the correlation between colour and
morphology indicating the crucial advantage of HST in distinguishing
morphologically between ellipticals and early spirals$^{79}$. Although
the samples remains small, field ellipticals do seem to be a less
homogeneous population than the clustered counterparts$^{80}$.  A more
powerful test of the merger hypothesis would be a direct estimate of
the comoving number density of evolved (red) ellipticals as a function
of redshift. Early results suggest a significant paucity of luminous
red objects at faint magnitudes$^{40,81,49}$.

A more complex topic is the history of disk formation. Gas cooling
onto pre-existing dark matter halos produces rotating disks but
star formation must be delayed until late times in order to avoid
an over-production of small spirals which would merge too readily.
Early formation also leads to a tidal transfer of angular momentum 
from the baryonic gas to the outer dark halo. To match the observed 
properties of local spirals, gas cooling has to be delayed to 
$z\simeq$1-2$^{82}$. Again, recent formation is preferred and strong
evolutionary effects are predicted.

However, the deepest HST images show no obvious decline with redshift 
in the abundance of spirals; well-formed examples are found at
$z\simeq$1$^{30,31}$. Unfortunately, a detailed census is not trivial to 
construct because evolutionary effects in stellar luminosity and disk 
scale size both affect the selection process. Surface photometry parameters 
such as the bulge/disk ratio, disk scale lengths and central 
surface brightnesses can be extracted from HST images$^{83}$ and these 
indicate large well-formed spirals were as abundant at $z\simeq$0.8 as 
they are today$^{31}$ with only modest changes in surface brightnesses 
over 0.3$<z<$1. A more detailed population analysis of HDF spirals 
further supports the suggestion of well-established stellar populations 
at $z\simeq$1-1.5$^{80}$.

Ultimately one wishes to characterise evolution in terms
of the assembling mass. The Keck telescope has provided the first 
rotation curves for HST-selected spirals to $z\simeq$1$^{84,85}$ and 
such dynamical data could be combined with independent photometric 
estimates of luminosity changes to yield mass growth rates for direct
comparison with theoretical models. At present, such analyses
are confused by the possibility of different selection criteria
and it has been argued that both datasets are consistent with
significant recent growth$^{86}$.

Ellipticals and spirals account for only half of the global evolutionary
trend between 0$<z<$1$^{41,42,56}$. The remainder arises from smaller,
more numerous, star-forming galaxies of irregular morphology$^{27-29}$
whose role remains unclear$^{19}$. As their fraction increases with
redshift, it is tempting to postulate these are ancestors of more regular
systems. Numerical simulations in which the gas is clumped make this an
attractive hypothesis$^{87}$. However, so many irregular sources are
seen, even at $z<$1, this seems unlikely.  Their existence alongside
well-formed spirals and ellipticals is evidence for continued star
formation governed by processes in addition to gravitational collapse
of gas into dark matter potentials$^{88}$.

\smallskip
{\bf Where Next?}
\medskip

HST has revolutionised this field by providing resolved images of faint
galaxies. Allied ground-based spectroscopy has not only provided
distances and luminosities, but long-slit techniques have supplied the
first rotation curves at high z. A new generation of
actively-controlled large telescopes are nearing completion and their
image quality at near infrared wavelengths will approach that of
HST$^{89}$. {\it Integral field} spectrographs$^{90}$, where the
sampling of a contiguous area of sky is reformatted to fit a long-slit
instrument, will considerably extend the capabilities. For example,
high redshift irregulars imply an erratic star formation history
consistent with a time sequence of bursts associated with each physical
component$^{91}$ (Figure 4). Are these merging systems in a chaotic
dynamical field or clumps in a well-established disk? Resolved 2-D
spectroscopy will lead to an explosion of new data on the dynamics,
excitation and dust content of young galaxies similar to that opened up
with HST itself.  The excitement generated by this year's SCUBA results
shows how dangerous it may be to rely on a narrow wavelength range.
More extensive sub-mm and mm surveys including those with the proposed
new millimetre arrays will provide more detailed information on the
nature and extent of star formation in distant sources.

Our first glimpse of the history of galaxies to redshifts $z\simeq$5
leads to the tantalising question of what happened before. The optimum
strategy for probing the ``dark age'' beyond depends on the amount of
dust at early times. If, as the limited SCUBA data currently suggests,
dusty sources share a similar redshift-dependence to those probed by
young stars then the early Universe may be relatively dust-free and
searches based on emission lines from neutral hydrogen will be
promising$^{92}$.  Such territory will be explored with projected
facilities such as the Next Generation Space Telescope and large
ground-based radio arrays.

\bigskip
\parskip=0pt

\pp 1. Hubble, E.  A spiral nebula as a stellar system: Messier 31.
Astrophys. J. {\bf 69}, 103-157 (1929)

\pp 2. Hubble, E. Extragalactic nebulae. Astrophys. J. {\bf 64}, 321-69 (1926)

\pp 3. Struck-Marcell, C. \& Tinsley, B.M. Star formation rates and
infrared radiation. Astrophys. J. {\bf 221}, 562-66 (1978)

\pp 4. Silk, J. \& Wyse, R.A. Galaxy formation and the Hubble Sequence.
Physics Reports {\bf 231}, 293-365  (1993)

\pp 5. Frenk, C.S., Ellis, R.S., Shanks, T., Heavens, A. \& Peacock, J.A.
Epoch of galaxy formation, Kluwer (1988)

\pp 6. Thompson, D. \& Djorgovski, S. Serendipitous long-Slit surveys
for primaeval galaxies. Astron. J. {\bf 110}, 982-994 (1995)

\pp 7. Toomre, A.  Mergers and some consequences, in `Evolution of 
Galaxies and Stellar Populations', eds. Tinsley, B.M. \& Larson, R.B. 
Yale University Observatory, New Haven, Conn., pp401-416 (1977)

\pp 8. Frenk, C.S., White, S.D.M., Efstathiou, G., Davis, M. 
Cold dark matter, the structure of galactic haloes and the origin
of the Hubble sequence. Nature, {\bf 317}, 595-597 (1985)

\pp 9. Jenkins, A. et al Evolution of structure in cold dark matter
universes. Astrophys. J. {\bf 499}, 20-40 (1998).

\pp 10. Kauffmann, G., White, S.D.M., Guiderdoni, B. The formation and 
evolution of galaxies within merging dark matter haloes. Mon. Not. R. 
Astron. Soc. {\bf 264}, 201-218 (1993)

\pp 11. Cole, S., Aragon-Salamanca, A., Frenk, C.S., Navarro, J.F.
\& Zepf, S. A recipe for galaxy formation. Mon. Not. R. Astron. Soc. 
{\bf 271}, 781-806 (1994)

\pp 12. Blumenthal, G., Faber, S.M., Primack, J.R. \& Rees, M.J. 
Formation of galaxies and large-scale structure with cold dark matter.
Nature {\bf 311}, 517-525 (1984)

\pp 13. Rees, M.J. \& Ostriker, J.P. Cooling, dynamics and fragmentation
of massive gas clouds - Clues to the masses and radii of galaxies
and clusters. Mon. Not. R. Astron. Soc. {\bf 179}, 541-559 (1977)

\pp 14. Fall, S.M. \& Rees, M.J. A theory for the origin of globular
clusters. Astrophys. J. {\bf 298}, 18-26 (1985)

\pp 15. White, S.D.M. \& Frenk, C.S. Galaxy Formation through hierarchical
clustering. Astrophys. J. {\bf 379}, 52-79 (1991)

\pp 16. Lilly, S.J., Cowie, L.L. \& Gardner, J.P. A deep imaging and
spectroscopic survey of faint galaxies. Astrophys. J. {\bf 369}, 79-105 (1991)

\pp 17. Metcalfe, N., Shanks, T., Fong, R. \& Roche, N. Galaxy number
counts - III. Deep CCD observations to B=27.5 mag. Mon. Not. R. Astron. 
Soc. {\bf 273}, 257-276 (1995)

\pp 18. Williams, R.E. et al The Hubble Deep Field: Observations, data
reduction and galaxy photometry. Astron. J. {\bf 112}, 1335-89 (1996)

\pp 19. Ellis, R.S. Faint blue galaxies. Ann Rev Astron Astrophys, {\bf 35},
389-443 (1997)

\pp 20. Gardner, J., Cowie, L., Wainscoat, R. Galaxy number counts from
K=10 to K=23. Astrophys. J. {\bf 415}, L9-12 (1993)

\pp 21. Djorgovski, S. et al Deep galaxy counts in the K-band with the 
Keck telescope. Astrophys. J. {\bf 438}, L13-16 (1995)

\pp 22. Lilly, S.J., Tresse, L., Hammer, F., Crampton, D., LeFevre, O.
The Canada-France Redshift Survey. VI. Evolution of the galaxy luminosity
function to z$\simeq$1. Astrophys. J. {\bf 455}, 108-24 (1995)

\pp 23. Ellis, R.S., Colless, M., Broadhurst, T.J., Heyl, J.S. \&
Glazebrook, K. Autofib Redshift Survey - I. Evolution of the galaxy
luminosity function. Mon. Not. R. Astron. Soc. {\bf 280}, 235-251 (1996)

\pp 24. Cowie, L., Songaila, A., Hu, E. \& Cohen, J. New insight on galaxy
formation and evolution from Keck spectroscopy of the Hawaii deep fields.
Astron. J. {\bf 112}, 839-64 (1996)

\pp 25. Cowie, L., Hu, E., Songaila, A. \& Egami, E. The evolution of the
distribution of star formation rates in galaxies. Astrophys. J. {\bf 481}, 
L9-13 (1997)

\pp 26. Munn, J.A. et al The Kitt Peak galaxy redshift survey and multicolor
photometry: Basic data. Astrophys. J. Suppl. {\bf 109}, 45-77 (1997)

\pp 27. Glazebrook, K., Ellis, R.S., Santiago, B. \& Griffiths, R.
The morphological identification of the rapidly-evolving population of
faint galaxies. Mon. Not. R. Astron. Soc. {\bf 275}, L19-22 (1995)

\pp 28. Driver, S.P., et al The morphological mix of field galaxies to I=24.25
($b_J\simeq$26) from a deep Hubble Space Telescope WFPC2 image. 
Astrophys. J. {\bf 449}, L23-27 (1995)

\pp 29. Cowie, L., Hu, E. \& Songaila, A. Detection of massive forming
galaxies at redshifts z$>$1. Nature {\bf 377}, 603-605 (1995) 

\pp 30. Brinchmann, J. et al Hubble Space Telescope imaging of the CFRS
and LDSS redshift surveys. I. Morphological Properties. Astrophys. J. 
{\bf 499}, 112-133 (1998)

\pp 31. Lilly, S.J. et al  Hubble Space Telescope Imaging of the CFRS
and LDSS Redshift Surveys. II. Structural parameters and the evolution of
disk galaxies to z$\simeq$1. Astrophys. J., {\bf 500}, 75-94  (1998)

\pp 32. Steidel, C. \& Hamilton, D. Deep imaging of high redshift QSO
fields below the Lyman limit. I - The field of Q0000-263 and galaxies
at z=3.4. Astron. J. {\bf 104}, 941-49 (1992)

\pp 33. Guhathakurta, P, Tyson, A.J. \& Majewski, S. A redshift limit for 
the faint blue galaxy population from deep U band imaging. Astrophys. J. 
{\bf 357}, L9-12 (1990)

\pp 34. Steidel, C., Giavalisco, M., Dickinson, M., Adelberger K.
Spectroscopy of Lyman break galaxies in the Hubble Deep Field.
Astron. J. {\bf 112}, 352-58 (1996)

\pp 35. Steidel, C., Giavalisco, M., Pettini, M., Dickinson, M. Adelberger, K. 
Spectroscopic confirmation of a population of normal star-forming galaxies 
at redshifts z$>$3. Astrophys. J. {\bf 462}, L17-21 (1996)

\pp 36. Lowenthal, J., Koo, D.C., Guzman, R., Gallego, J. et al.,
Keck spectroscopy of redshift z$\simeq$3 galaxies in the Hubble Deep Field.
Astrophys. J. {\bf 481}, 673-688 (1997)

\pp 37. Pettini, M. et al The spectra of star forming galaxies at high
redshift in `The Ultraviolet Universe at Low and High Redshift: Probing
the Progress of Galaxy Evolution' ed. Waller, W. et al AIP Conf. Proc.
408, 279-289 (1997)

\pp 38. Giavalisco, M., Steidel, C., Macchetto, D. Hubble Space Telescope
imaging of star-forming galaxies at redshifts z$>$3. Astrophys. J. 
{\bf 470}, 189-94 (1996).

\pp 39. Adelberger, K. et al A counts-in-cells analysis of Lyman break
galaxies at z$\sim$3. Astrophys. J. in press (astro-ph/9804236) 

\pp 40. Dickinson, M. Color-selected high redshift galaxies and the
HDF, in `Hubble Deep Field' eds. Livio, M. et al, CUP, pp219-244 (1998).

\pp 41. Madau, P. et al High-redshift galaxies in the Hubble Deep Field:
colour selection and star formation history to z$\sim$4. Mon. Not. R. 
Astron. Soc. {\bf 283}, 1388-1404 (1996)

\pp 42. Connolly, A., Szalay, A.S., Dickinson, M., SubbaRao, M., Brunner R.
The Evolution of the global star formation history as measured from the
Hubble Deep Field. Astrophys. J. {\bf 486}, L11-14 (1997).

\pp 43. Baugh, C., Cole, S., Frenk, C.S. \& Lacey, C.G. The epoch of
galaxy formation. Astrophys. J. {\bf 498}, 504-521 (1998).

\pp 44. Governato, F. et al The seeds of rich galaxy clusters in
the universe. Nature {\bf 392}, 359-361 (1998)

\pp 45. Roche, N. Ratnatunga, K., Griffiths, R.E., Im, M. \& Naim, A.
Galaxy surface brightness and size evolution to z$\sim$4. Mon. Not. R. 
Astron. Soc. {\bf 293}, 157-176 (1998)

\pp 46. Bouwens, R., Broadhurst, T.J. \& Silk J. 
Cloning Hubble Deep Fields: A model-independent measure of galaxy evolution.
Astrophys. J. in press (astro-ph/9710291)

\pp 47. Abraham, R.G., et al Galaxy morphology to I=25 in the Hubble Deep
Field. Mon. Not. R. Astron. Soc. {\bf 279}, L47-52 (1996)

\pp 48. Broadhurst, T.J., Ellis, R.S. \& Glazebrook, K. Faint galaxies -
evolution \& cosmic curvature. Nature {\bf 355}, 55-58 (1992)

\pp 49. Kauffmann, G. \& Charlot, S. The K-band luminosity function at
z=1: a powerful constraint on galaxy formation theory. Mon. Not. R. Astron.
Soc. {\bf 297}, L23-28 (1998)

\pp 50. Fall, S.M., Pei, Y.C. \& Charlot, S.
Cosmic emissivity and background intensity from damped Lyman-alpha
galaxies. Astrophys. J. {\bf 464}, L43-46 (1996)

\pp 51. Pettini, M., Smith, L.J., King, D. \& Hunstead, R.W. 
The Metallicity of high-redshift galaxies: the abundance of Zinc
in 34 damped Lyman alpha systems from z=0.7 t0 3.4. Astrophys. J.
{\bf 468}, 665-680 (1997)

\pp 52. Lu, L., Sargent, W.L.W. \& Barlow, T.A. The N/Si abundance ratio
in 15 damped Lyman alpha galaxies: Implications for the origin of
Nitrogen. Astron. J. {\bf 115}, 55-61 (1998)

\pp 53. Songaila, A. A lower limit to the universal density of metals
at z$\simeq$3. Astrophys. J. {\bf 490}, L1-4 (1997)  

\pp 54. Storrie-Lombardi, L., McMahon, R.G. \& Irwin, M.J.  Evolution
of neutral gas at high redshift: implications for the epoch of galaxy
formation. Mon. Not. R. Astro. Soc. {\bf 283}, L79-83 (1996)

\pp 55. Meurer, G., Heckman, T.M., Lehnert, M.D., Leitherer, C. \&
Lowenthal, J. The panchromatic starburst intensity limit at high redshift.
Astron. J. {\bf 114}, 54-68 (1997) 

\pp 56. Madau, P., Pozzetti, L., Dickinson, M. The star formation history
of field galaxies. Astrophys. J. {\bf 498}, 106-116 (1998)

\pp 57. Sanders, D.B., Mirabel, I.F. Luminous infrared galaxies.
Ann. Rev. Astron. Astrophys. {\bf 34}, 749-792 (1996)

\pp 58. Mann, R.G. et al Observations of the Hubble Deep Field with the
Infrared Space Observatory - IV. Association of sources with Hubble Deep
Field galaxies. Mon. Not. R. Astron. Soc. {\bf 289}, 482-489 (1997)

\pp 59. Blain, A.W., Longair, M.S. Observing strategies for blank-field
surveys in the submillimetry waveband. Mon. Not. R. Astron. Soc. {\bf 279},
847-858 (1996)

\pp 60. Smail, I., Ivison, R.G. \& Blain, A.W. A deep Sub-millimeter
survey of lensing clusers: a new window on galaxy formation and evolution.
Astrophys. J. {\bf 490}, L5-8 (1997)

\pp 61. Hughes, D., Serjeant, S., Dunlop, J., Rowan-Robinson, M. et al
Unveiling dust-enshrouded star formation in the early Universe: a sub-mm
survey of the Hubble Deep Field. Nature {\bf 394}, 241-247 (1998)
 
\pp 62. Barger, A., Cowie, L.L., Sanders, D.B., Fulton, E. et al 
Dusty star forming galaxies at high redshift. Nature {\bf 394}, 248-251 (1998)

\pp 63. Ivison, R. et al A hyperluminous galaxy at z=2.8 found in a deep
submillimetre survey. Mon. Not. R. Astron. Soc., in press (astro-ph/9712161)

\pp 64. Baugh, C., Cole, S. \& Frenk, C.S. Evolution of the Hubble sequence
in hierarchical models for galaxy formation. Mon. Not. R. Astron. Soc. 
{\bf 283}, 1361-1378 (1996)

\pp 65. Dressler, A. Galaxy morphology in rich clusters - implications for the
formation and evolution of galaxies. Astrophys. J. {\bf 236}, 351-365 (1980)

\pp 66. Butcher, H. \& Oemler, A. The evolution of galaxies in clusters. 
I - ISIT photometry of Cl0024+1654 and 3C295. Astrophys. J. {\bf 219}, 
18-30 (1978)

\pp 67. Dressler, A. et al Evolution since z=0.5 of the morphology-density
relation for clusters of galaxies. Astrophys. J. {\bf 490}, 577-591 (1997)

\pp 68. Smail, I. et al A catalog of morphological types in 10 distant
rich clusters of galaxies. Astrophys. J. Supp. {\bf 110}, 213-225 (1997)

\pp 69. Moore, B., Katz, N., Lake, G., Dressler, A. \& Oemler, A.
Galaxy harassment and the evolution of clusters of galaxies. Nature 
{\bf 379}, 613-616 (1996)

\pp 70. Gunn, J.E. \& Gott, J.R. On the infall of matter into clusters
of galaxies and some effects on their evolution. Astrophys. J. {\bf 176}, 
1-19 (1972)

\pp 71. Mihos, C. Morphology of galaxy mergers at intermediate redshift.
Astrophys. J. {\bf 438}, L75-78 (1995)

\pp 72. van Dokkum, P. et al The color-magnitude relation in Cl1358+62 at
z=0.33: Evidence for significant evolution in the S0 population.
Astrophys. J. 500, 714- (1998)

\pp 73. Morris, S. et al Galaxy evolution in the z=0.4274 cluster
MS1621.5+2640. Astrophys. J. in press (astro-ph/9805216))

\pp 74. Baade, W. Galaxies and their stellar populations in 
`Stellar Populations' ed. O'Connell, D. (Vatican Obs) p3-24 (1957)

\pp 75. Bower, R.G., Lucey, J.R., Ellis, R.S.  Precision photometry
of early type galaxies in the Coma and Virgo clusters - a test of
the universality of the colour/magnitude relation. II - Analysis.
Mon. Not. R. Astron. Soc. {\bf 254}, 601-613 (1992)

\pp 76. Ellis, R.S. et al The homogeneity of spheroidal populations 
in distant clusters. Astrophys. J. {\bf 483}, 582-596 (1997)

\pp 77. Kauffmann, G., Charlot, S. \& White, S.D.M. 
Detection of strong evolution in the population of early-type galaxies.
Mon. Not. R. Astron. Soc. {\bf 283}, L117-22 (1996)

\pp 78. Kauffmann, G. Hierarchical clustering and the Butcher-Oemler
effect. Mon. Not. R. Astron. Soc. {\bf 274}, 153-160 (1995)

\pp 79. Schade, D. et al Hubble Space Telescope Imaging of the CFRS
and LDSS Redshift Surveys. III - Properties of field ellipticals
with 0.2$<$z$<$1. Astrophys. J. (submitted)

\pp 80. Abraham, R.G., Ellis, R.S., Fabian, A.C., Tanvir, N.R. \&
Glazebrook, K. The star formation history of the Hubble sequence:
Spatially resolved colour distributions of intermediate redshift
galaxies in the Hubble Deep Field. Mon. Not. R. Astron. Soc. (in press)

\pp 81. Zepf, S.E. Formation of elliptical galaxies at moderate redshifts.
Nature {\bf 390}, 377-379 (1997)

\pp 82. Weil, M., Efstathiou, G. \& Eke, V. R. The formation of disk
galaxies. Astrophys. J. in press (astro-ph/9802311)

\pp 83. Schade, D. et al Canada-France Redshift Survey: Hubble Space
Telescope imaging of high-redshift field galaxies. Astrophys. J. 
{\bf 451}, L1-4 (1995)

\pp 84. Vogt, N. et al Optical rotation curves of distant field galaxies:
results at redshifts to z$\simeq$1. Astrophys. J. {\bf 465}, L15-18 (1996)

\pp 85. Vogt, N. et al Optical rotation curves of distant field galaxies:
Sub-L$^{\ast}$ systems. Astrophys. J. {\bf 479}, L121-124 (1997)

\pp 86. Mao, S., White, S.D.M. \& Mo, H-J. The Evolution of Galactic 
Disks. Mon. Not. R. Astron. Soc. in press (astro-ph/9712167)

\pp 87. Noguchi, M. Clumpy star-forming regions as the origin of the
peculiar morphology of high-redshift galaxies. Nature {\bf 392}, 254-256 (1998)

\pp 88. Babul, A. \& Rees, M.J. On dwarf elliptical galaxies and the
faint blue counts. Mon. Not. R. Astron. Soc. {\bf 255}, 346-350 (1992)

\pp 89. Giacconi, R. First Light of the VLT Unit Telescope 1, ESO 
Messenger, No. 92, p2 (1998)

\pp 90. van der Riest, C., Bacon, R., Georgelin, Y., Le Coarer \&
Monnet, G. Astronomical uses of integral field spectroscopy: 
present applications at CFHT and future developments in `Instrumentation
in Astronomy VIII, eds. Crawford, D. \& Craine, E.R., Proc. SPIE {\bf 2198}, 1376-1384 (1994).

\pp 91. Ellis, R.S. HDF: Introduction and motivation in `The Hubble Deep 
Field', ed. Livio, M. et al, CUP, p27-38 (1998).

\pp 92. Hu, E., Cowie, L.L. \&  McMahon, R.G. The density of Lyman-alpha
emitters at very high redshift. Astrophys. J. {\bf 502}, L99-103 (1998).

\medskip
{\bf Acknowledgements:} I acknowledge numerous discussions with my 
colleagues at Cambridge and Carnegie Observatories and all my 
collaborators on the Morphs, CFRS/LDSS redshift survey and Hubble 
Deep Field programmes.

\vfill\eject
\bigskip
\centerline{\bf Figure Captions}
\bigskip

{\bf Figure 1:} [Main panel] The Hubble Deep Field: the deepest 
exposure undertaken at optical wavelengths$^{18}$ and an inspiration 
to other instruments which have followed suit and imaged this 
undistinguished 2 $\times$ 2 arcmin patch of sky in Ursa Major 
(see panel below). Despite the considerable exposure time, the rate 
of increase in galaxy surface density with optical faintness appears 
to decline beyond a certain level (inset) suggesting optical and
infrared images reach to epochs before most galaxies had formed their 
stars. [Bottom panels] Whether the HST image offers a representative 
view of the deep Universe is a hotly contested issue. The same field 
studied with different instruments can address this question although 
none can yet match its depth and resolution. From left to right: 
ground-based near-infrared (Dickinson, Kitt Peak), mid-infrared 
(Infrared Space Observatory$^{58}$, sub-mm (SCUBA, James Clerk Maxwell Telescope$^{61}$). 

[Inset: Galaxy number magnitude counts - a simple but powerful
cosmological tool (see ref 19 for a recent summary). The logarithmic 
slope of the count-magnitude relation is a measure of the rate of evolution 
of the population with look-back time. The steep slope at moderate 
magnitudes contrast with a slightly flatter one at fainter limits. Galaxies 
near the point of inflexion contribute most to the integrated background 
light of the night sky.]

\bigskip
{\bf Figure 2:} (Centre panel) The history of star formation inferred 
from optical and near-infrared ground-based surveys$^{31,41,42}$. Below 
a redshift $z\simeq$1, magnitude-limited surveys define the luminosity density
by direct census. Above $z$=2, the Lyman-limit imaging technique 
(left inset) locates those sources which reveal the characteristic
signature of a `drop-out' in flux in a short wavelength filter$^{32,33}$.
At intermediate redshifts, redshifts are estimated by fitting the
spectral energy distribution defined by 4-7 colours$^{42}$ (right inset).
It is argued that the empirical data validates an early prediction of
late galaxy formation central to the cold dark matter model of structure
formation$^{15,43}$. The key issue is whether the optical data includes
most of the flux from star forming galaxies.

[Insets: (left) Finding distant galaxies by the Lyman-limit `drop out'
technique (after Dickinson$^{40}$). Absorption by neutral hydrogen
produces a characteristic signature at 912 \AA\ which, when redshifted,
extinguishes a galaxy in a short wavelength filter. By locating that
wavelength where the galaxy is no longer visible, an approximate
redshift can be obtained. The technique is useful for sources with
redshifts $z>$2.3$^{35}$. (right) Photometrically-determined redshifts:
By fitting the spectral energy distribution of a galaxy observed in
many filters, the approximate redshift can be found$^{42}$. The technique
remains controversial since it is largely untested against the 
spectroscopic values (abscissa) in the region of interest 1.2$<z<$2.5]

\bigskip
{\bf Figure 3:} Nature versus nurture: did galaxies transform from
one morphological type to another and if so, what physical processes
were involved? (left panel) Local spirals and ellipticals
defined according to Hubble's original scheme$^{2}$ contrasted
with images of fainter, more distant galaxies taken with HST. A
surprising fraction of high redshift galaxies appear disturbed and irregular
indicating they are being seen in a juvenile state$^{30,47}$. (right panel) 
The morphology-density relation; clustered regions today contain many
ellipticals and lenticulars (S0s) but relatively few spirals$^{65}$. 
Does this trend reflect primordial conditions or recent environmental
evolution? Systematic studies of clusters at earlier times$^{66-68}$
have tracked the evolution in this relationship confirming that
environmental effects shape the present-day appearance of many galaxies.

\bigskip
{\bf Figure 4:} Determining the physical state of a young galaxy:
Beyond $z\simeq$1 a typical galaxy is physically small 
consisting of numerous knots of activity. Does this represent chaotic 
activity in a well-established system$^{87}$ or are we witnessing the 
active merger of numerous components as implied in hierarchical models?
Integral field spectroscopy$^{90}$ (top panel) offers the opportunity of
sampling the dynamical and astrophysical properties of each component
of a young galaxy. This can be coupled with detailed stellar population
analyses based on multi-colour HST data which age-date 
physically-distinct sub-units$^{80,91}$ (bottom panel). Such techniques 
will be exploited for the brightest galaxies with 1$<z<$3 using the 
new generation of large telescopes, but such studies of more distant 
sources, including those in the `dark age' beyond $z>$5 await 
the Next Generation Space Telescope.

\bye